\documentclass[aps,pra,longbibliography,showkeys,showpacs,groupedaddress,twocolumn,10pt]{revtex4-1}
\usepackage{amssymb}
\usepackage{graphicx}
\usepackage{dcolumn}
\usepackage{bm}
\usepackage{amsmath}
\usepackage{subfigure}
\usepackage{float}
\usepackage{color}
\usepackage{float}
\usepackage{hhline}

\begin{document}
\title{Observation of many-body dynamics of Rydberg atoms based on antiblockade using two-color excitation}
\author{Suying Bai$^{1,4}$}
\author{Xuedong Tian$^{2}$}
\author{Xiaoxuan Han$^{1,4}$}
\author{Yuechun Jiao$^{1,4}$}
\author{Jinhui Wu$^{3}$}
\thanks{Corresponding author: jhwu@nenu.edu.cn}
\author{Jianming Zhao$^{1,4}$}
\thanks{Corresponding author: zhaojm@sxu.edu.cn}
\author{Suotang Jia$^{1,4}$}
\affiliation{$^{1}$State Key Laboratory of Quantum Optics and Quantum Optics Devices, Institute of Laser Spectroscopy, Shanxi University, Taiyuan 030006, China}
\affiliation{$^{2}$College of Physics Science and Technology, Guangxi Normal University, Guilin 541004, China}
\affiliation{$^{3}$School of Physics, Northeast Normal University, Changchun 130024, China}
\affiliation{$^{4}$Collaborative Innovation Center of Extreme Optics, Shanxi University, Taiyuan 030006, China}
\date{\today}

\begin{abstract}

We investigate a long-range interaction between $64D_{5/2}$ Rydberg-atom pairs and antiblockade effect employing a two-color excitation scheme. The first color (pulse A) is set to resonantly excite the Rydberg transition and prepare a few seed atoms, which establish a blockade region due to strong long-range interaction between Rydberg-atom pairs. The second color (pulse B) is blue detuned relative to Rydberg transition and enables further Rydberg excitation of atoms by counteracting the blockade effect.
It is found that a few seed atoms lead to a huge difference of the Rydberg excitation with pulse B.
The dynamic evolution of antiblockade excitation by varying the pulse-B duration at 30-MHz blue detuning is also investigated. The evolution result reveals that a small amount of seed atoms can trigger an avalanche Rydberg excitation.
A modified superatom model is used to simulate the antiblockade effect and relevant dynamic evolution. The simulations are consistent with the experimental measurements.
\end{abstract}
\maketitle

\section{Introduction}
Creation of ultracold gases in $\thicksim \mu$K regime has opened a new avenue for
investigating interacting many-body systems. For nondegenerate gases such as
atoms in a magneto-optical trap (MOT), the interaction between ground state atoms is very weak. However, Rydberg atoms, atoms with principal quantum numbers $n \gtrsim$ 10, can strongly interact, even in a dilute gas, due to their strong long-range dipole-dipole ($\thicksim n^4$) and Van der Waals (vdw) interaction ($\thicksim n^{11}$)~\cite{Gallagher}. These scalings allow for accurately controlling the interactions over a huge range by varying $n$. Ultracold Rydberg gases are analogue to amorphous solids because an atom moves only about 3\% of the average interatomic spacing during $\thicksim$ 1~$\mu$s time scale of typical experimental interest and thus can be deemed stationary to a good approximation, which make Rydberg atoms idea candidate for investigating the many-body dynamics~\cite{Ates2007}.

The strong interaction between Rydberg atoms shifts the energy level and suppresses the further excitation of the neighborhood, leading to a blockade effect~\cite{Tong2004,singer2004}. The blockade radius $R_B$ is usually defined as the distance between two atoms at which the Rydberg interaction energy equals the excitation linewidth ($\gamma_e$), e.g. $V_{int}(R_B)$ = $\gamma_e$. Within the region of $R < R_B$, only one Rydberg atom can be excited, yielding thus the blocked volume of a single Rydbeg excitation, i.e., a superatom (SA) formed by a Rydberg atom in superposition with many ground-state atoms~\cite{Liu,Fleischhauer}.
Based on the blockade effect, Rydberg atoms can be employed to implement quantum information processing~\cite{Saffman,Lukin}, quantum registers~\cite{Endres,Xia,Kim,Lester}, single-photon sources~\cite{Dudin2012,Peyronel} and transistors~\cite{Li2014,Gorniaczy2014,Tiarks2014}. On the other hand,
Rydberg ensembles with strong interaction are also ideal for investigating excitation transfer~\cite{Gunter,Barredo,BarredoLabuhn,Maineult}, antiblockade effect~\cite{Letscher2017}, and many-body dynamics~\cite{Ates2007}. Antiblockade effect means, in particular, that two or more Rydberg atoms can be excited in a blockade region under appropriate driving schemes~\cite{Ates}.

In this work, we investigate the long-range interaction between 64$D_{5/2}$-atom pairs and antiblockade effect  in a dilute sample of cold cesium atoms. The Rydberg excitation in the blockade regime is found to be largely enhanced with a two-color double-resonance method, where pulse-A  excites resonantly one seed atom, whereas the frequency of pulse-B is detuned for compensating the interaction-induced shift of nearby atoms to allow further Rydberg excitation. Many-body dynamics of the Rydberg excitation is also studied by varying the interaction time of pulse-B. Our experimental observations are consistent with relevant theoretical simulations based on a modified SA model, in which the interaction-induced shift is assumed to be large but finite.

\section{Long-range interaction model}
We first introduce the theoretical model for calculation of the Rydberg-pair interaction potential. The details are discussed in our previous work~\cite{Han,Han2019}, so we briefly describe the model below. For calculating the interaction of a Rydberg-atom pair, we consider two $nD_J$ Rydberg atoms, denoted $a$ and $b$, with an interatomic separation $\bf{R}$.
To simplify the calculation, the quantization axis and $\bf{R}$ are both chosen along the $z$-axis, see Fig.~1(a).
The relative positions of the Rydberg electrons are ${\bf{r}}_a$ and ${\bf{r}}_b$. The interatomic distance $R$ is larger than the LeRoy radius~\cite{Le Roy}, $R_{LR}$. The Hamiltonian of the Rydberg-atom pair is written as:
\begin{eqnarray}
\hat{H} = \hat{H}_a + \hat{H}_b + \hat{V}_{int},
\end{eqnarray}
where $\hat{H}_{a(b)}$ is the Hamiltonian of atom $a(b)$, and $\hat{V}_{int}$ denotes the multipole interaction between the Rydberg-atom pair. $\hat{V}_{int}$ is taken as~\cite{Schwettmann2006,Deiglmayr2014,Han,Han2019}

\begin{figure}[ht]
\vspace{-1ex}
\centering
\includegraphics[width=0.4\textwidth]{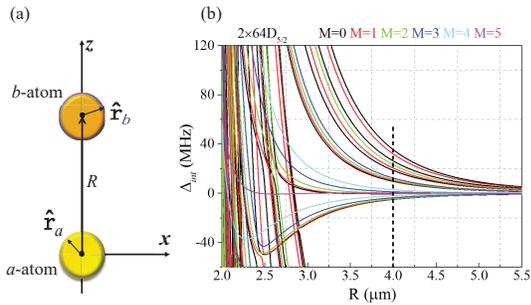}
\vspace{-1ex}
\caption{(Color online) (a) Two-atom system. Rydberg atoms $a$ and $b$, separated by $R > R_{LR}$, are placed on the $z$-axis, ${\bf{r}}_{a}$ and ${\bf{r}}_{b}$ are the relative positions of the Rydberg electrons in atoms $a$ and $b$. (b) Calculations of potential curves for cesium $64D_{5/2}$ Rydberg-atom pair, with indicated $M$, maximal order $q_{max}$ = 6 and energy defect 30~GHz. The setting of the basis size and single-atom orbital angular momentum see text. A vertical dashed line indicates atom-pair interactions at $R = 4~\mu$m. }
\end{figure}

\begin{eqnarray}
\widehat{V}_{int}&=& \sum_{q=2}^{q_{max}} \frac{1}{R^{q+1}} \sum_{\substack{L_{a}=1 \\ L_b=q-L_a}}^{q_{max}-1}
\sum_{\Omega=-L_{<}}^{L_{<}}f_{ab\Omega}
\hat{Q}_{a} \hat{Q}_{b} \\
f_{ab\Omega}&=&\frac{(-1)^{L_{b}} \, (L_{a}+L_{b})!}{\sqrt{(L_{a}+\Omega)!(L_{a}-\Omega)!(L_{b}+\Omega)!(L_{b}-\Omega)!}}
\end{eqnarray}
where $L_{a(b)}$ are the multipole orders of atoms $a(b)$, and the $L_{<}$ is the lesser of $L_{a}$ and $L_{b}$. The sum over $q$=$L_{a}$+$L_{b}$ starts at 2, because atoms are neutral and have no monopole moment, and is truncated at a maximal order $q_{max}$. The factor $f_{ab\Omega}$ depends on $L_{a}$, $L_{b}$ and the counting index $\Omega$ under the third sum. The $\hat{Q}_{a(b)}$ =
$\sqrt{\frac{4\pi}{2L_{a(b)}+1}}\widehat{r}_{a(b)}^{L_{a(b)}}Y_{L_{a(b)}}^{\Omega}(\widehat{\bf{r}}_{a(b)})$, where the single-atom operators $\hat{{\bf{r}}}_{a(b)}$ is the relative position of the Rydberg electron in atom $a(b)$ and operators $\hat{Q}_{a(b)}$ include radial matrix elements, $\hat{r}_{a(b)}^{L_{a(b)}}$, and spherical harmonics that depend on the angular parts of the Rydberg-electron positions, $Y_{L_{a(b)}}^{\pm \Omega}(\hat{{\bf{r}}}_{a(b)})$.

We diagonalize the Hamiltonian of the Rydberg-atom pair on a dense grid of the internuclear separation, $R$.
The $R$ range is 2.0~$\mu$m to 5.5~$\mu$m and the number of radial steps is 400. To improve the quality of plots of the potential curves at small $R$, the radial steps are chosen equidistant in $R^{-3}$. Because of global azimuthal symmetry, the projection of the sum of the electronic angular momenta, $M=m_{Ja} + m_{Jb}$, is conserved.
In Fig.~1(b), we present the calculation of the interaction potential curves with the single-atom orbital angular momentum space $\ell \leq \ell_{max}$ and $m_{J} \leq m_{Jmax}$, and the two-body states of energy defects 30~GHz and the range of effective principal quantum numbers ${\rm{int}}(n_{\rm eff0})- \delta < n_{\rm eff} < {\rm{int}} (n_{\rm eff0})+\delta +1$, ($n_{\rm eff0}$ is the effective quantum number of the Rydberg state). ${\rm{int}} (n_{\rm eff0})$ denotes the integer part of $n_{\rm eff0}$, and $\delta$ is a parameter for the principal quantum number range.

For every molecular state $M$, only one potential curve exhibits minima that could give rise to bound Rydberg macrodimer~\cite{Han}, while all other potential curves show repulsive interaction. For the typical MOT density $N \sim 10^{10}$ ~cm$^{-3}$, we have the average interatomic separation $R \sim $ 4.0~$\mu$m, marked with dashed line in Fig.~1(b). The repulsive interaction between Rydberg 64$D_{5/2}$ atom pair $\Delta_{int}=\langle \widehat{V}_{int} \rangle$ being 2$\pi \times$ 5-40~MHz. It is this strong long-range repulsive interaction that leads to the blockade effect. In following section, we focus on the blue-detuned side of the $64D_{5/2}$ atomic resonance and investigate the antiblockade phenomenon with a two-color excitation proposal.

\section{Experimental setup}

\begin{figure}[ht]
\vspace{-1ex}
\centering
\includegraphics[width=0.4\textwidth]{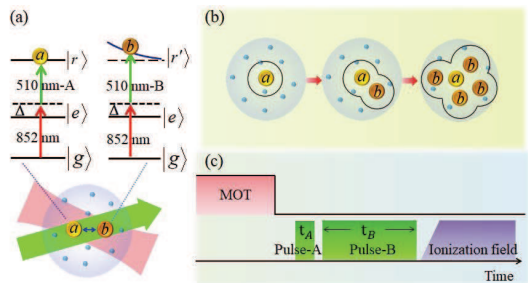}
\vspace{-1ex}
\caption{(Color online) (a) Schematic of experiments and relative level diagram for the two-color excitation. The 852-nm and 510-nm excitation beams cross over with the angle about 40$^{\circ}$ at the MOT center. The pulse A resonantly excites Rydberg atom (atom $a$). For studying antiblockade effects, the frequency of the 510-nm component of pulse B is scanned to blue-detuned side to facilitate the excitation of Rydberg atoms (atom $b$) in blockade region. $\Delta$ = 360~MHz is the single-photon detuning of pulse A and B. (b) Sketch of the antiblockade effect and dynamic evolution of Rydberg excitation. Black lines display the blockade regions. When the pulse-B detuning compensates the shift due to the Rydberg interaction, an atom $b$ in blockade region is excited, that would trigger an avalanche excitation of the Rydberg atoms. (c) Timing sequence. After switching off MOT beams, we sequentially apply pulses A and B. Prepared Rydberg atoms are detected using the electric-field ionization method. Variations of the pulse-B duration, $t_B$, allow us to study the evolution of Rydberg-atom excitation and many-body dynamics.
}
\end{figure}

Antiblockade experiments are performed in a standard cesium MOT with a temperature $T \sim$~100~$\mu$K and a density $N \sim$~$~10^{10}~$~cm$^{-3}$. The atomic density is controlled by delaying the Rydberg excitation beams.
After switching off the MOT beams, we successively apply pulses A and B, as sketched in Fig.~2(a) and a timing sequence in Fig.~2(c).
Both pulses A and B are in two-photon excitation scheme. The lower-transition laser (852~nm, Toptica DLpro, $\sim~100$~kHz linewidth) is stabilized to the $|6S_{1/2}, $F$ = 4\rangle$ ($|g\rangle$) $\to$ $|6P_{3/2},  $F'$ = 5 \rangle$ ($|e\rangle$) transition using a polarization spectroscopy technique~\cite{Pearman}, and is shifted off-resonance from intermediate level $|e\rangle$ by 360~MHz using a double-pass acousto-optic modulator (AOM).
The upper-transition laser (510~nm, Toptica TA SHG110, $\sim$1~MHz linewidth) is stabilized to a 64$D_{5/2}$ Rydberg transition using a F-P cavity with a finesse 15000 and double-passed through another AOM. For the pulse pair A, the 510-nm AOM frequency is set to resonantly excite Rydberg atoms ($a$-atoms). For the subsequent pulse B, the 510-nm AOM frequency is blue detuned to maximum 100~MHz.
During the scan, the B-pulse laser power is held fixed using a PID (proportional integral derivative) feedback loop that controls the RF power supplied to the 510-nm AOM. The 852-nm laser has a power of $\thicksim$270~$\mu$W and Gaussian waist of $\omega_{852}$ $\simeq$ 180~$\mu$m, whereas
510-nm beam has a waist $\omega_{510}$ $\simeq$ 40~$\mu$m at the MOT center.

The 852-nm and 510-nm excitation beams cross over with an angle about 40$^{\circ}$ at MOT center yielding an elliptical excitation region. This setup allow the atomic density within the excitation region maintain nearly same.
The excitation region is surrounded by three pairs of field-compensation electrodes, which allow us to reduce stray electric fields to less than 50~mV/cm via Stark spectroscopy.
Rydberg atoms are detected using the electric-field ionization method (ionization ramp rise time 3~$\mu$s). The extracted ions are detected with a microchannel plate (MCP) detector.

Figure 2(b) demonstrates the basic idea of the antiblockade excitation. Once pulse A resonantly prepares Rydberg-atom $a$, the neighbor atoms in blockade region, black line in the left part of Fig.~2(b), will undergo level shifts due to strong Rydberg interactions as shown in Fig.~1(b). When pulse-B detuning approaches to this level shift, the neighbor atom $b$ is excited to Rydberg atom, see middle part of Fig.~2(b). The Rydberg atom $b$ is thought as a new seed atom extending the blockade region. This process allows the pulse B to excite further Rydberg atoms in blockade region that would trigger an avalanche excitation of the Rydberg atoms, as shown in the right part of Fig.~2(b).

\section{Experimental observation of antiblockade}
\begin{figure}[tbp]
\vspace{1ex}
\centering
\includegraphics[width=0.4\textwidth]{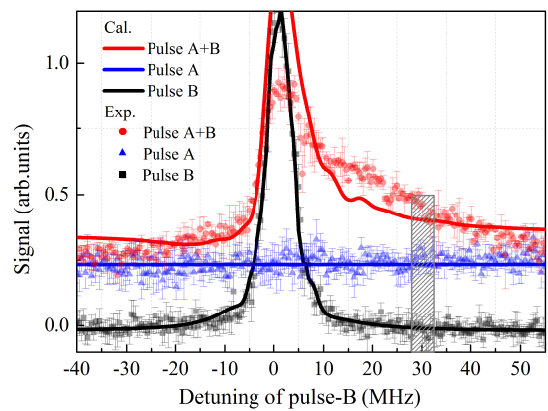}
\vspace{-1ex}
\caption{(Color online) Measurements (samples) and calculations with a modified SA model (lines) of the Rydberg signal versus the detuning of the pulse B for switching on only pulse A (blue triangles), pulse B (black squares) and both A and B (red circles). For the case of only pulse A switched on, the signal of Rydberg atoms remain unchanged as pulse A is set to resonant with the Rydberg transition. For the case of only pulse B on, Rydberg atoms can be excited only when the frequency of the pulse B is scanned to resonance with the $64D_{5/2}$ Rydberg transition. Whereas the case of both pulses on, Rydberg excitation are facilitated at large blue detuned range of the pulse B. A small amount of seed atoms can enlarge the Rydberg excitation at blue detuned side of pulse B. A gray square denotes the 30~MHz detuning point, where the Rydberg signal is factor of 2 larger than that of pulse A.}
\end{figure}

We firstly study how a few seed atoms excited by the pulse A affect the excitation of pulse B. We do the test by switching on the pulse A or B and both.
Figure~3 presents the Rydberg excitation spectra as a function of pulse B detuning with the pulse A duration 0.4 $\mu$s, and pulse B 6.0 $\mu$s, respectively. The two pulses are suitably arranged such that pulse A prepares a few seed Rydberg atoms, as seen in Fig.~3. Rydberg-atom number excited with pulse A is about 5 factor smaller than with pulse B at resonance. The Rydberg signal excited with only pulse-A keep unchanged as we scan pulse-B frequency. For the only pulse-B case, analogue to normally two-photon Rydberg excitation, atoms can be excited to Rydberg state at 0 detuning with a linewidth $\backsimeq$ 8~MHz, extracted with Lorentz fitting to the data.
When the pulse-B detuning is larger than the linewidth of Rydberg spectrum, there is no any atom can be excited to Rydberg state, shown with the black squares in Fig.~3.

However, when we apply a 0.4-$\mu$s pulse A before the pulse B, Rydberg excitation can be enhanced at large detuning range, (10-50~MHz here, depending on the atomic density and Rydberg state selected). For example, the Rydberg excitation is zero at 20-MHz blue detuned side when only pulse B is applied, but a factor of 2.5 larger when we apply pulse A before pulse B. At 30-MHz detuned point, as marked with the gray square in Fig.~3, the enhanced Rydberg excitation is less than the 20-MHz-detuning case, it is a factor of 2 larger.
This demonstrates that the existence of a small amount of seed atoms $a$ greatly facilitates the excitation probability of pulse B at blue detuned side. We attribute this phenomenon to an antiblockade effect.

For further investigating the dynamics of Rydberg excitation, we vary the pulse-B duration and do a series of measurements while keeping the detuning of the pulse A and B fixed. We set pulse A on resonance and pulse B 30~MHz blue detuned to the Rydberg transition. In Fig.~4, we present the Rydberg excitation as a function of the pulse-B duration. For the only pulse A or B switching on, the Rydberg populations do not change as increasing pulse B duration, as denoted with black squares for pulse B on and blue triangles for pulse A on in Fig.~4. Pulse A excites a few seed atoms that has minor variation due to the fluctuation of laser power. Pulse B excites nothing as the frequency is detuned, 30~MHz, far from resonant of Rydberg transition. However, for the case of both pulses on, as displayed with red circles in Fig.~4, Rydberg excitation demonstrates completely different characteristics. Rydberg populations display minor increasing for the pulse-B duration less than 2~$\mu$s, then fast increasing as pulse B duration and approaches to saturation as pulse B up to $t_B \thicksim$ 5~$\mu$s.

\begin{figure}[tbp]
\vspace{1ex}
\centering
\includegraphics[width=0.4\textwidth]{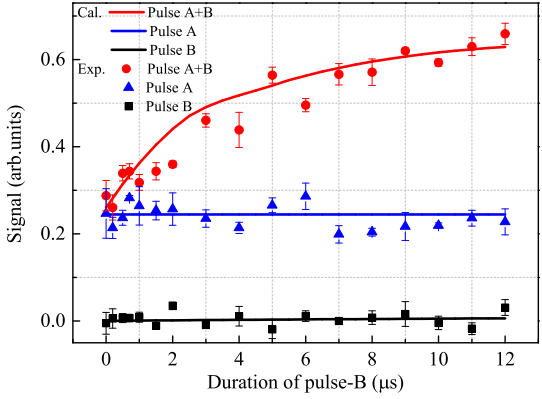}
\vspace{-1ex}
\caption{(Color online) Measurements (samples) and simulations (lines) of Rydberg excitations as a function of the pulse B duration while keeping the pulse B 30~MHz blue detuning and pulse A resonance to the $64D_{5/2}$ transition. Three cases are for switching on only pulse A (blue triangles), pulse B (black squares) and both A and B (red circles). The increase of pulse B duration dose not affect the Rydberg excitation for the cases of switching on pulse A or B only, whereas strongly modify the case of both pulses on. Rydberg excitation displays minor increasing with the pulse-B duration at $t_B$ $\lesssim$ 2~$\mu$s and fast increasing at $\thicksim 5~\mu$s and then approaches to saturation.}
\end{figure}

\section{Simulations based on modified superatom model}

In order to explain the observations above, we use a modified SA model to simulate the experiments based on the following two
considerations. First, the lower (852~nm) and upper (510~nm) excitation lasers have Rabi frequencies $\Omega _{852}^{j}=2\pi \times 48.5$~MHz, $\Omega
_{510}^{j}=2\pi \times 6.4$~MHz, and single-photon detuning $\Delta$ = $\Delta _{852}^{j}\approx -\Delta
_{510}^{j}\approx 2\pi \times 360$~MHz for pulse $j\in \{A,B\}$, so we can adiabatically eliminate intermediate level $\left\vert e\right\rangle $
to have an effective Rabi frequency $\Omega _{eff}^{j}=\Omega _{852}^{j}\Omega
_{510}^{j}/2\Delta^{j}\approx~2\pi \times 0.43$~MHz that couples levels $\left\vert
g\right\rangle $ and $\left\vert r\right\rangle $. Second, the average
atomic distance is $R \sim 4.0$~$\mu $m from the
averaged atomic density $N \sim 10^{10}$ $cm^{-3}$, so we know from Fig. 1(b) that
the multipole interaction between two nearest-neighbor
(next-nearest-neighbor atoms) atoms is much stronger (smaller) than $\Omega
_{eff}^{a}$. With these considerations, we
assume that there are only two strongly interacting Rydberg atoms in each SA, and
the multipole interactions between the atomic pairs in different SAs are negligble. Aiming
to attain more accurate results, we write down the effective
Hamiltonian without setting $\Delta _{int}\rightarrow \infty $
and without neglecting level $\left\vert e\right\rangle $,

\begin{widetext}
\begin{eqnarray*}
H_{eff}^{j} &=&\sum\nolimits_{i=a,b}\left[ \Delta _{852}^{j}\left\vert
e_{i}\right\rangle \left\langle e_{i}\right\vert +(\Delta _{852}^{j}+\Delta
_{510}^{j})\left\vert r_{i}\right\rangle \left\langle r_{i}\right\vert \right]
+\Delta _{int}\left\vert r_{a}r_{b}\right\rangle \left\langle
r_{b}r_{a}\right\vert  \\
&&+\sum\nolimits_{i=a,b}\left[ \Omega _{852}^{j}\left\vert e_{i}\right\rangle
\left\langle g_{i}\right\vert +\Omega _{510}^{j}\left\vert r_{i}\right\rangle
\left\langle e_{i}\right\vert +\Omega _{852}^{j}\left\vert g_{i}\right\rangle
\left\langle e_{i}\right\vert +\Omega _{510}^{j}\left\vert e_{i}\right\rangle
\left\langle r_{i}\right\vert \right],
\end{eqnarray*}%
\end{widetext}
for atoms $a$ and $b$ in each SA by assuming $\hbar =1$. Taking $H_{eff}^{j}$
into the master equation $\partial _{t}\rho ^{j}=-i[H_{eff}^{j},\rho ^{j}]$
of density operator $\rho ^{j}$ and adding phenomenally the decay ($\Gamma
_{\mu \nu }$) and dephasing ($\gamma _{\mu \nu }$) rates, we obtain a
set of dynamical equations for $9\times 9$ density matrix elements $\rho
_{\mu \nu ,\mu \nu }^{j}$ with the former (latter) $\mu $ and $\nu $
referring to the three levels of atom $a$ ($b$) in the excitation
process of pules $j$. Then we use this set of equations to simulate the $A$ ($B$)
excitation process by applying the resonant (detuned) pulse $A$ ($B$%
) with $\Delta _{int}=\{-4.0\sim 34.0\}\times 2\pi $~MHz, taken from Fig.~1(b), for $21$ interaction potentials
formed with different magnetic sublevels with $M\in \{0,1,2,3,4,5\}$ of weight factor $\zeta\in
\{6,5,4,3,2,1\}$. We further note that $(i.)$ pulse A has a fixed duration
$t_{A}=0.4$ $\mu $s while the duration $t_{B}$ for pulse B will be varied
in the range $\{0\sim 12\}$~$\mu $s in our simulations; $(ii.)$ our
numerical results are attained as a sum of $21$ independent numerical
realizations with different magnetic sublevels serving as level $\left\vert
r_{i}\right\rangle $; ($iii.$) the two-photon resonant A excitation
process typically results in a blockade effect while the two-photon
off-resonant B excitation process is expected to yield an antiblockade
effect.

Simulations of the excitation spectra are displayed with solid lines in Fig.~3 for three indicated excitation conditions. It is seen that the calculations reproduce the spectra well for pulse A or B on. Whereas for the case of both pulses on, the calculated excitation spectrum is just basically consistent with the experimental results. To be more concrete, the calculations agree well at pulse B detuning $\lesssim$ 15~MHz and $\gtrsim$ 30~MHz. However, at range $\thicksim$ 15-30~MHz, the calculated Rydberg excitation is less than measurements, which may be attributed to the insufficiency of a two-body SA model we used.

In Fig.~4, we plot the simulations on dynamics of Rydberg excitation for the fixed pulse B detuning at 30~MHz, marked with the gray square in Fig.~3.
The calculation reproduces the experimental measurements very well again. The agreement between calculations and measurements for only with pulse A(B) on illustrates that our model is valid. For the case of both pulses on, simulations demonstrate faster increase than the measurements at $t_B$ $<$ 3~$\mu$s.
Increasing speed of the population becomes slower at $t_B$ $\thicksim 3~\mu$s and approaches maximum showing saturation behavior.
The minor difference between calculations and measurements at $t_B \backsimeq$ 1-3~$\mu$s is attributed to the following two reasons. First reason is related to the interaction term we used in the model. For D-type Rydberg state, the interaction potentials are very complicated as seen in Fig.~1. For example, a Rydberg-atom pair in the M=0 case has six potential energy curves. On the other hand, as mentioned above, we consider only two-body interaction for simplicity, which may be not enough to answer for the interplay of different potential energy curves.

\section{Conclusion}
In summary, we have investigated the long-range interaction between $64D_{5/2}$ Rydberg-atom pairs that yields an excitation blockade phenomenon. Two-color excitation scheme is used to realize the Rydberg excitation of those atoms that are blocked in the blockade region, e.g. antiblockade effect. The first color (pulse A) is resonant to the Rydberg transition for preparing a few seed atoms, yielding a blockade sphere. The second color (pulse B) is blue detuned relative to Rydberg resonance for facilitating the Rydberg excitation. It is found that a few seed atoms lead to a huge difference of the Rydberg excitation of pulse B. Pulse B excites ground state to Rydberg state at 0-detuning without seed atoms, but can enable Rydberg excitation at much large detuning range with a few seed atoms induced by pulse A. The pulse B enhanced excitation depends on its frequency detuning. The dynamic evolution of antiblockade excitation by varying the pulse B duration reveals that a small amount of seed atoms can trigger an avalanche Rydberg excitation.

A modified SA model has been adopted to well simulate the antiblockade effect and dynamic evolution. The small deviation between simulation and measurement can be attributed to two facts: (i.) interaction potentials of D-type (64$D_{5/2}$) Rydberg state are complicated including many repulsive potential curves and avoided crossings; (ii.) we consider only two-body interaction for simplicity, which is not accurate enough as cold Rydberg gases are analogue to amorphous solids showing many-body behavior~\cite{Ates2007}. In future, we will use S-type Rydberg state with simpler interaction potential and higher principal quantum number to study the antiblockade excitation dynamics both experimentally and theoretically.

We thank G. Raithel for discussions on the calculation of Rydberg-atom pairs potential. The work was supported by the National Key R$\&$D Program of China (Grant No. 2017YFA0304203), the National Natural Science Foundation of China (Grants Nos. 61675123, 61775124 and 11804202), the State Key Program of National Natural Science of China (Grant No. 11434007 and 61835007), and Changjiang Scholars and Innovative Research Team in University of Ministry of Education of China (Grant No. IRT\_17R70). X.T. thanks the support of National Natural Science Foundation of China (Grant No. 11847018)

\end{document}